\documentclass[prl,aps,twocolumn]{revtex4-1}
\usepackage{mathbbold}
\usepackage{amsfonts}
\usepackage{amsmath}
\usepackage{graphicx}
\usepackage{color}
\usepackage{bm}
\usepackage{subfig}
\begin{document}
\title{Fast detection and automatic parameter estimation of a gravitational wave signal with a novel method}
\author{Yan Wang}
\affiliation{Max-Planck-Institut f\"ur Gravitationsphysik (Albert-Einstein-Institut), Callinstra{\ss}e 38, 30167 Hannover, Germany}
\email{yan.wang@aei.mpg.de} \pacs{}
\begin{abstract}
The detection of gravitational wave usually requires to match the measurement data with a large number of templates, which is computationally
very expensive. Compressed sensing methods allow one to match the data with a small number of templates and interpolate the rest. However,
the interpolation process is still computationally expensive. In this article, we designed a novel method that only requires to match
the data with a few templates, yet without needing any interpolation process. The algorithm worked well for signals with relatively high SNRs.
It also showed promise for low SNRs signals.
\end{abstract}
\maketitle
%%%%%%%%%%%%%%%%%%%%%%%%%%%%%%%%%%%%%%%%%%%%%%%%%%%%%%%%%%

\emph{Introduction}--While gravitational wave (GW) signals contain invaluable physical information, extracting this information from the noisy data is quite challenging. Most of the time, GW signals are weaker than the instrumental noise at any instant, but they are predictable and long lived \cite{Sathyaprakash09}. This gives a way to build up signal-to-noise ratio (SNR) over time by tracking the signals coherently with matched filtering \cite{Jaranowski12}. However, this requires the templates to be exactly the same as the true signal to recover the optimal SNR, or at least resemble the true signal sufficiently in order not to lose much SNR \cite{Owen96}. Since the template waveforms depend on several parameters, one needs to match the data with a huge number of templates in the high dimensional parameter space. Therefore, a normal grid-based search is usually computationally extremely expensive, or even prohibitive. The reduction of the computational cost lies in the center of the modern GW data analysis.

There are several categories of algorithms, successfully reducing the computational cost, such as reduced bases (RB) \cite{Field11}, singular value decomposition (SVD) \cite{Cannon10} and principal component analysis (PCA) \cite{Heng09}. These methods make use of the fact that each template is strongly correlated with the templates in its neighbourhood in the parameter space. Therefore, its SNR can be effectively interpolated from the SNRs of the templates in its neighbourhood. In other words, the likelihood surface on the grid of the template bank has special properties (sparsity), which allows the compressed sensing \cite{Candes06} algorithms to apply. Instead of using all the templates in the bank, one only needs to calculate the SNRs of a few so-called basis templates (which are different from the original templates), and then interpolate the SNRs of all the other templates in the bank. It is extremely fast to perform matched filtering on that few basis templates comparing to the original bank of templates. However, the interpolation (or sometimes referred to as the reconstruction) process is still computationally expensive.

We wish to design a novel method, which requires to perform matched filtering on a few templates, and in the meantime does not require any interpolation
stage (or can automatically reconstruct the parameters of the GW signal). However, this method currently requires a relatively high SNR of the signal. The detailed description of the method and the preliminary simulation results are shown in the following.

\emph{GW data analysis routine}--First of all, we briefly review the convention and notations of the GW data analysis. Usually, the measurement data can be expressed as $s = A h_* + n$, where $n$ is the noise, $A$ is the amplitude of the signal, $h_*$ is the normalized signal in the measurement, which satisfies $\langle h_*|h_* \rangle=1$. The inner product of two time
series $a(t)$ and $b(t)$ is defined as follows
\begin{eqnarray}
\langle a|b\rangle=\int_{-\infty}^{\infty}\frac{\tilde{a}^*(f)\tilde{b}(f)}{S_n(f)}df,
\end{eqnarray}
where $\tilde{a}(f),\tilde{b}(f)$ are the Fourier transforms of $a(t)$ and $b(t)$. $S_n(f)$ is the so-called two-sided
noise power spectral density (PSD), usually defined as $\textrm{E}[\tilde{n}^*(f')\tilde{n}(f)]=S_n(f)\delta(f-f')$.

The GW data analysis problem that we want to solve is formulated as follows. For a set of normalized candidate templates
$h_i=h(\Theta_i)$ (we choose the template index $i=1,\dots,2^N$ for convenience) characterized by parameters $\Theta_i$, we want to determine which one is present in the measurement, hence obtaining the parameters $\Theta_*$ of the signal. Notice that $\Theta$ denotes a set of waveform parameters. For clarity, we require the templates to be nearly independent $ \langle h_i|h_j \rangle \,\ll 1,\,  (i\neq j)$. This is not generally true for a whole template bank. However, one can easily divide the entire template bank into a group of smaller template banks, within which the templates are nearly independent.

We assume that the true signal $h_*$ belongs to the template family, $*\in \{1,2,\dots,2^N\}$. The inner product between the measurement data and a template is denoted as
\begin{eqnarray}
x_i &\equiv& \langle s|h_i \rangle \nonumber \\
&=& A \langle h_*|h_i \rangle + \langle n|h_i \rangle,
\end{eqnarray}
thus the expectation and the variance are
\begin{eqnarray}
\textrm{E}(x_i) &=& A \delta_{*,i} \\
\textrm{Var}(x_i) &=& \textrm{E}[\langle h_i|n \rangle \langle n|h_i \rangle] \nonumber \\
&=& \langle h_i | h_i \rangle = 1.
\end{eqnarray}
By identifying the largest inner product $x_*$, we can detect the signal $h_*$ and estimate its parameters $\Theta_*$. When the inner product $x_*$ is much larger
than its standard deviation $\sqrt{\textrm{Var}(x_*)}=1$, the significance is high. The above shows a normal search strategy, which requires to perform $2^N$ inner products.

\emph{The novel method}--In the following, we will describe a novel search algorithm. First, we express the waveform indices $i$ in binary, hence
each index is an $N$-digit binary number (e.g. $001011011\dots$). Then, we define $N$ sets $\mathcal{P}_k$ ($k=1,2,\dots,N$)
such that $\mathcal{P}_k$ consists of all the indices $i$ whose $k$-th digit is $1$. A new template family is defined based
on these sets
\begin{eqnarray}
H_k = \sum_{i\in \mathcal{P}_k} h_i.
\end{eqnarray}
The inner products of these new templates with the measurement data are
\begin{eqnarray}
X_k &\equiv& \langle s|H_k \rangle \nonumber \\
&=& \sum_{i\in \mathcal{P}_k} \langle s|h_i \rangle.
\end{eqnarray}
The expectation of $X_k$ is
\begin{eqnarray}
\textrm{E}(X_k)=\left\{
\begin{aligned}
A,\;\;\;\;\;  *\in \mathcal{P}_k\\
0,\;\;\;\;\;  *\notin \mathcal{P}_k
\end{aligned}
\right.
\end{eqnarray}
The variance can be calculated as follows
\begin{eqnarray}
\textrm{Var}(X_k)&=& \textrm{E}[\langle n | \sum_{i\in \mathcal{P}_k} h_i \rangle^2] \nonumber \\
&=& \sum_{i,j\in \mathcal{P}_k} \langle h_i | h_j \rangle.
\end{eqnarray}
Since the templates $h_i$ are nearly independent, we have
\begin{eqnarray}
\textrm{Var}(X_k)&=& \sum_{i\in \mathcal{P}_k} \langle h_i | h_i \rangle \nonumber \\
&=& 2^{N-1}.
\end{eqnarray}
Suppose $*\in \mathcal{P}_a$ and $*\notin \mathcal{P}_b$, then
\begin{eqnarray}
\textrm{E}(X_a-X_b)&=& A \\
\textrm{Var}(X_a-X_b) &=& \textrm{E}[\langle n | \sum_{i\in \mathcal{P}_a} h_i - \sum_{j\in \mathcal{P}_b} h_j \rangle ] \nonumber \\
&=& \sum_{i\in \{\mathcal{P}_a\cup \mathcal{P}_b - \mathcal{P}_a \cap \mathcal{P}_b\}}  \langle h_i | h_i \rangle  \nonumber \\
&=& 2^{N-1}.
\end{eqnarray}
When the expectation $A$ is much larger than the standard deviation $2^{(N-1)/2}$, we can set
some threshold $\mathcal{T}$ between $A$ and $2^{(N-1)/2}$. Based on this threshold, a binary number can be obtained
as follows: if $X_k>\mathcal{T}$, the $k$-th bit of this binary number is $1$, otherwise its
$k$-th digit is set as $0$. This binary number can be converted to a decimal number $i_0$.
The method identifies the waveform $h_{i_0}$ with parameters $\Theta_{i_0}$ to be most probably present
in the data. In this new approach, we
have used $N$ templates instead of $2^N$ templates to detect the signal and estimate its parameters.
The computational cost is thus reduced from $\mathcal{C}\cdot 2^N$ to $\mathcal{C}\cdot N$. Notice that, if each
inner product of the data and a template provides one bit of information (above or below a certain threshold), $N$ is the minimum required number
of templates to distinguish $2^N$ sets of candidate parameters.

\emph{Simulation}--To exemplify the performance of the novel method, we consider the following chirp waveform family
\begin{eqnarray}
h(t;f,\dot{f}) = \mathcal{A}\cos (2\pi f t+ \pi \dot{f} t^2),
\end{eqnarray}
where $\mathcal{A}$ is the normalization constant, $f$ and $\dot{f}$ are the two intrinsic parameters to be estimated.
We have simulated $100$ seconds measurement data at $1\,$kHz with different SNRs. The parameters of the true signal are
$f_*=100\,$Hz and $\dot{f}_*=0.2\,$Hz/s. We have considered $2^6$ candidate waveforms with the parameter mesh grid
$$f=\{70,80,90,100,110,120,130,140\}\,\textrm{Hz},$$
$$\dot{f}=\{-0.3,-0.2,-0.1,0,0.1,0.2,0.3,0.4\}\,\textrm{Hz/s}.$$

\begin{figure}
\centering
\includegraphics[width=0.5\textwidth]{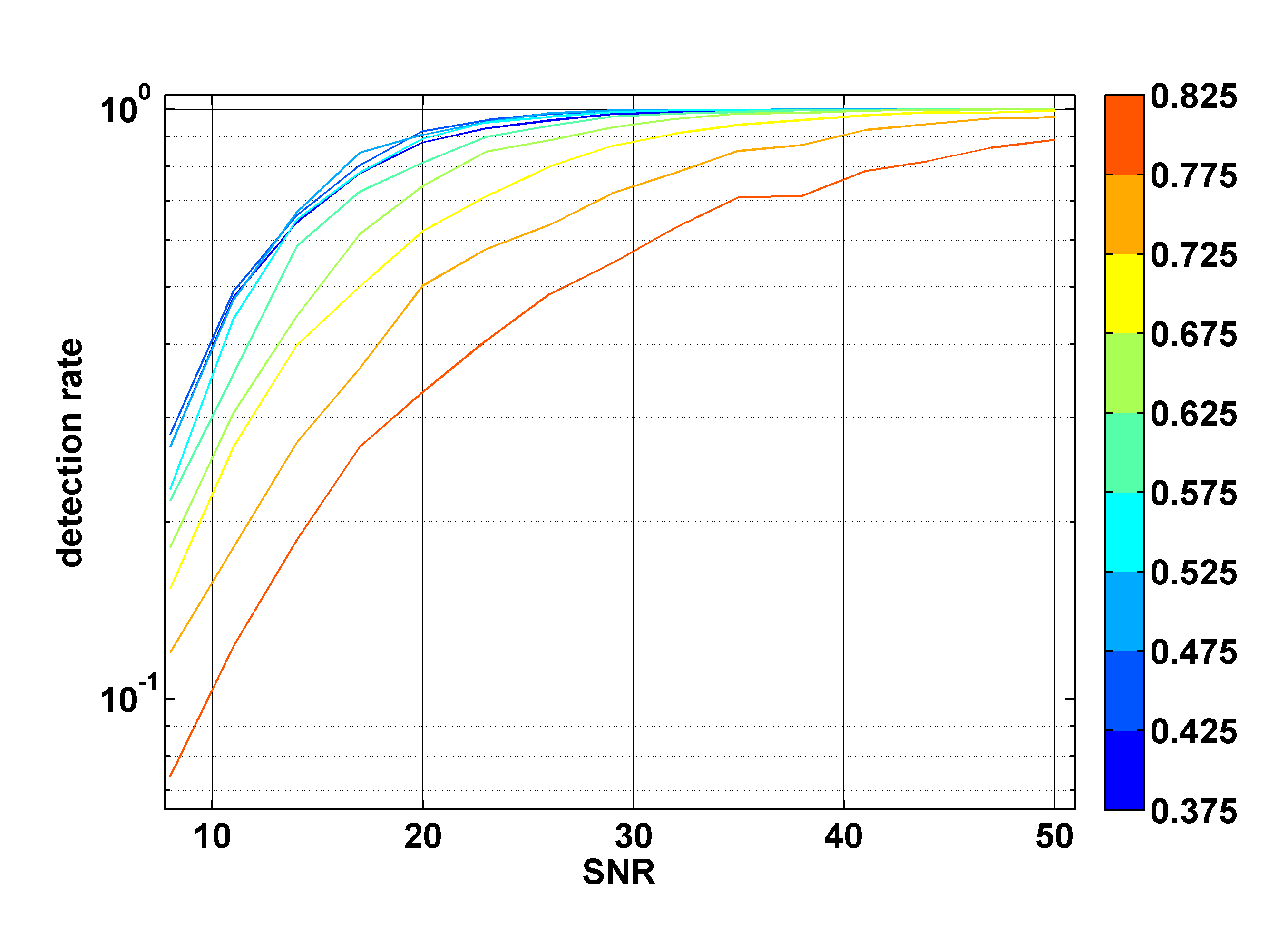}
\caption{ \label{fig:detectionRate} Detection rate at different SNRs and thresholds. The color bar indicates the value of the coefficient
$c$. The algorithm achieves the optimal performance, when $c$ is around $0.5$. The detection rate is above $99\%$, when SNR is above $30$.}
\end{figure}

The threshold is simply chosen as $\mathcal{T} = c\cdot\max(X_k)$, where we have tried several values of the coefficient $c$.
The SNR varies from $8$ to $50$ with a uniform spacing $3$. For each combination of SNR and the threshold, we
carried out a Monte Carlo simulation with $1000$ different noise realizations. If the algorithm identifies the true signal
and its true parameters, the detection is successful. The success rate is called the detection rate. Fig.~\ref{fig:detectionRate}
shows the detection rate at different SNRs and thresholds, where the color bar indicates the value of the coefficient
$c$. The best performance is realized by setting the coefficient $c$ around $0.5$. For signals with SNR higher than $30$,
the detection rate of the algorithm is above $99\%$. Thus, the algorithm with the least number of new templates works efficiently
at relatively hight SNRs. However, at low SNRs, the detection rate is low. We will see whether we could improve the detection
rate by slightly increase the computational cost.

\begin{figure}
\centering
\includegraphics[width=0.5\textwidth]{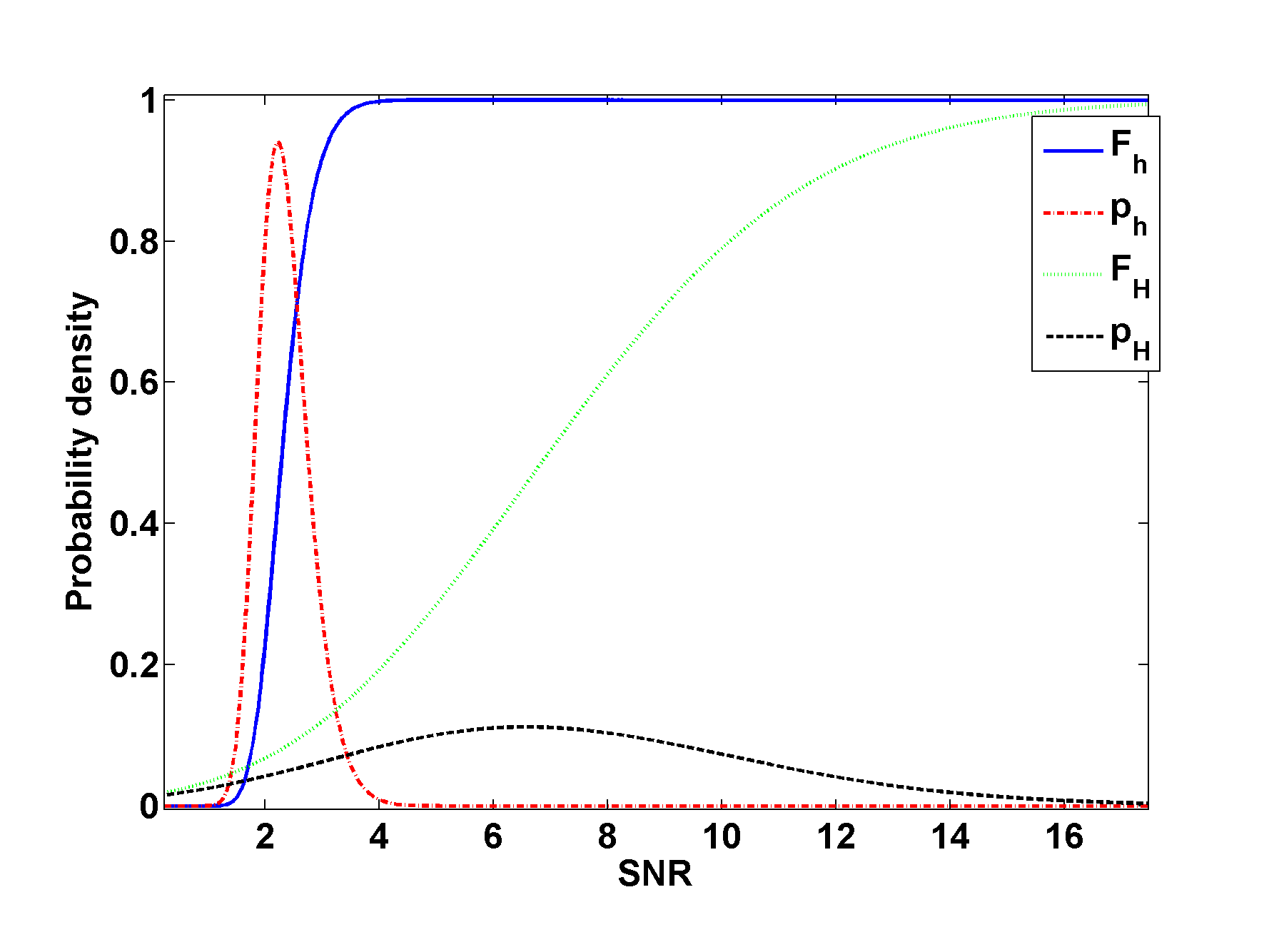}
\caption{ \label{fig:PDF} The probability density functions and cumulative distribution functions of the random part of $x_i$ and $X_k$,
which are $\langle n|h_i\rangle$ and $\langle n|H_k\rangle$.}
\end{figure}

\emph{Features of the algorithm}--For the set of $2^N$ independent templates $h_i$, if $2^N$ is smaller than the number of samples in the observation data,
$x_i=\langle s|h_i\rangle$ are also independent. To characterize the performance of the algorithms, we want to examine to what extent can the noise mimic
a signal. Since the signal part of $x_i$ only contributes a DC bias to its probability distribution, we can ignore the DC part and only consider the random part of
$x_i$, which is $\langle n|h_i \rangle$. It can be shown without much effort that the probability density function of the maximum of these $2^N$ random
variables $x_i$ is the following
\begin{eqnarray}
p_{\max}(x) = \frac{2^N}{\sqrt{2^{2^{N+1}-1}\pi}} \left[ 1 + \textrm{erf}\left( \frac{x}{\sqrt{2}} \right) \right]^{2^N-1}e^{-\frac{x^2}{2}}, \nonumber \\
\end{eqnarray}
where the error function
$\textrm{erf}(x)$ is defined as $\textrm{erf}(x)=\frac{2}{\sqrt{\pi}}\int_0^x e^{-x^2}dx$.
%Since these random variables follow Gaussian distribution with a zero mean, which is symmetric about the y-axis, the minimum of these random variables has a probability
%density function as follows
%\begin{eqnarray}
%p_{\min}(x) = \frac{2^N}{\sqrt{2^{2^{N+1}-1}\pi}} \left[ 1 + \textrm{erf}\left( \frac{-x}{\sqrt{2}} \right) \right]^{2^N-1}e^{-\frac{x^2}{2}}.
%\end{eqnarray}
Similarly, the probability density function of the maximum of the $N$ random
variables $X_k$ turns out to be the following
\begin{eqnarray}
p_{\max}(X) = \frac{N}{\sqrt{2^{3N-2}\pi}} \left[ 1 + \textrm{erf}\left( \frac{X}{\sqrt{2^N}} \right) \right]^{N-1}e^{-\frac{X^2}{2^N}}. \nonumber \\
\end{eqnarray}
For the case we considered, we have $N=6$. The probability density functions and cumulative distribution functions of the random part of $x_i$ and $X_k$
are shown in Fig.~\ref{fig:PDF}, which tells us how large SNRs could be mimicked by pure noise. As expected, in case of $X_k$, the noise could mimic larger
SNRs. This can also be seen from the larger standard deviation of $X_k$. In fact, this is the reason for the drop in the detection rate at low SNRs in Fig.~\ref{fig:detectionRate}.

\begin{figure}[htbp]
\centering
\subfloat[SNR=50]{
\begin{minipage}[t]{0.25\textwidth}
\centering
\includegraphics[width=1.0\textwidth]{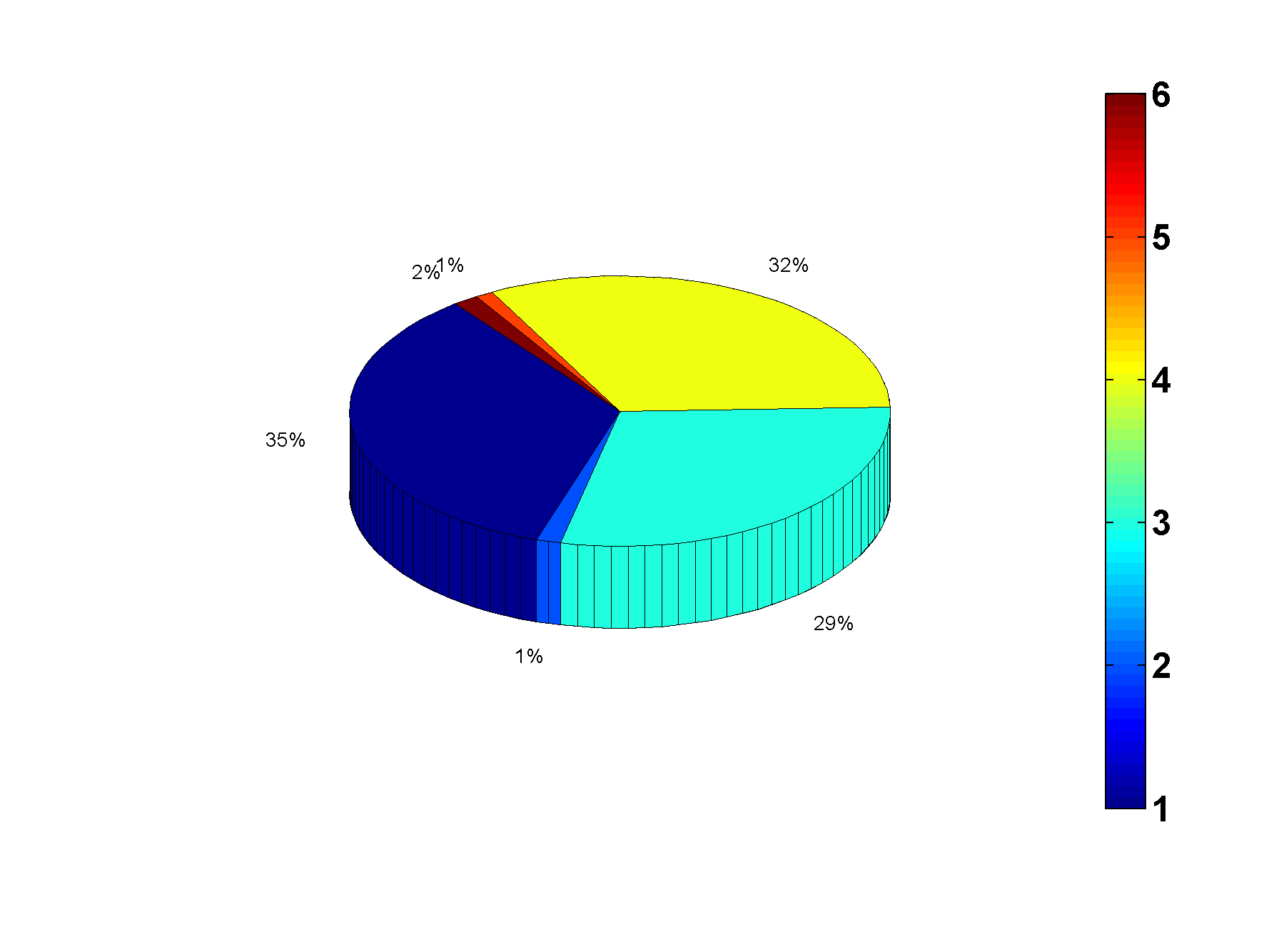}
\end{minipage}
}
\subfloat[SNR=40]{
\begin{minipage}[t]{0.25\textwidth}
\centering
\includegraphics[width=1.0\textwidth]{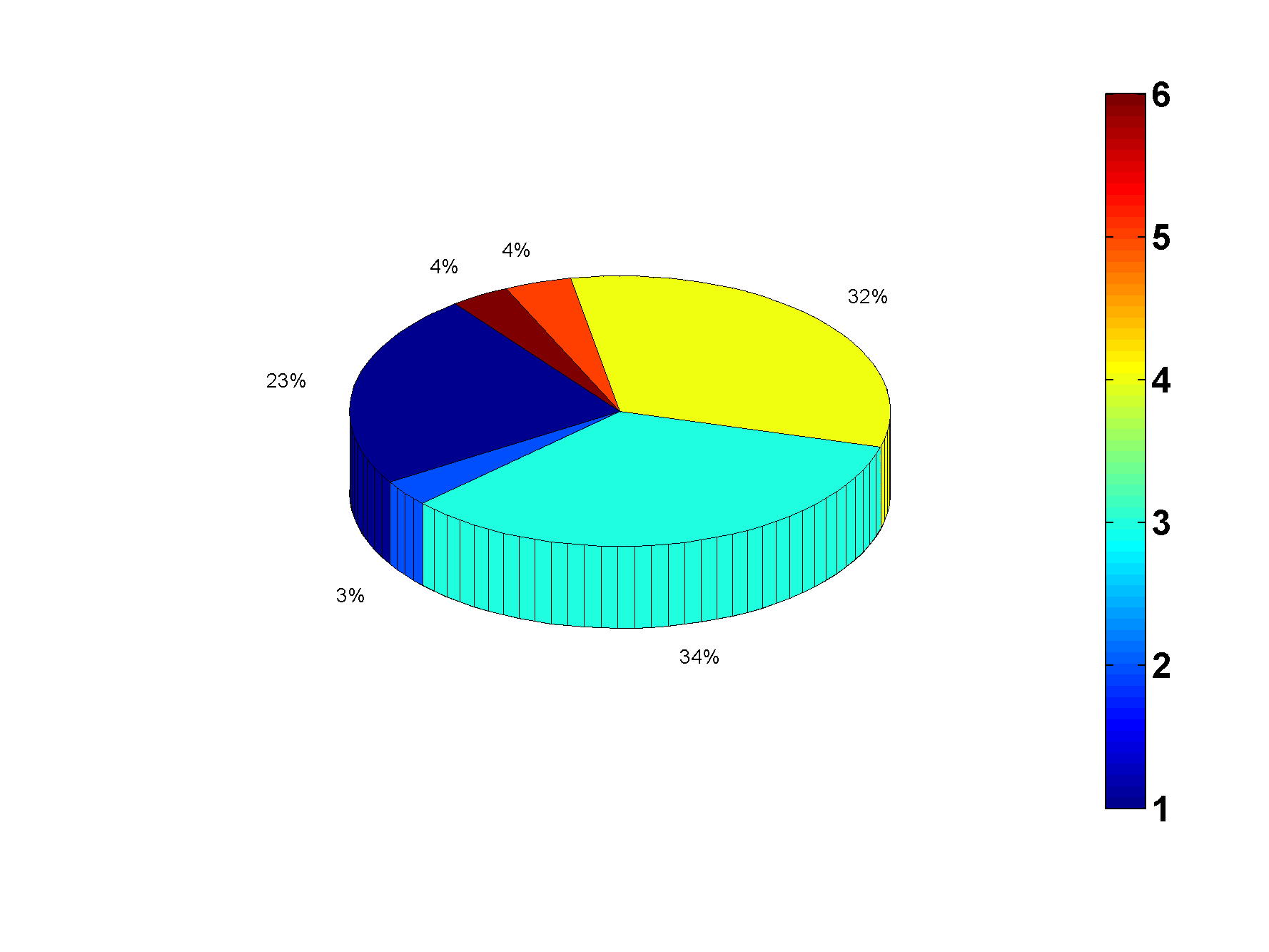}
\end{minipage}
}
\\
\subfloat[SNR=30]{
\begin{minipage}[t]{0.25\textwidth}
\centering
\includegraphics[width=1.0\textwidth]{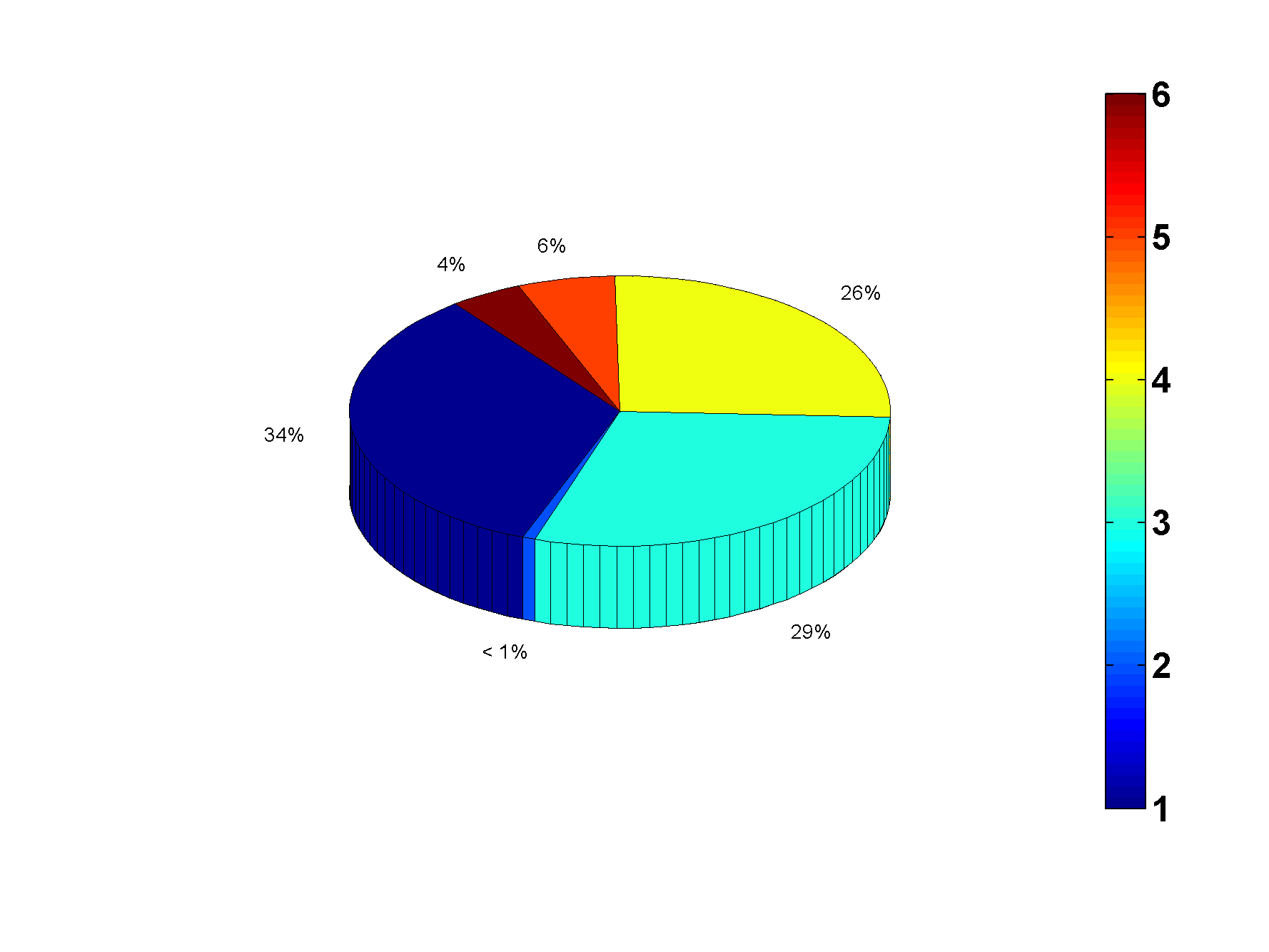}
\end{minipage}
}
\subfloat[SNR=20]{
\begin{minipage}[t]{0.25\textwidth}
\centering
\includegraphics[width=1.0\textwidth]{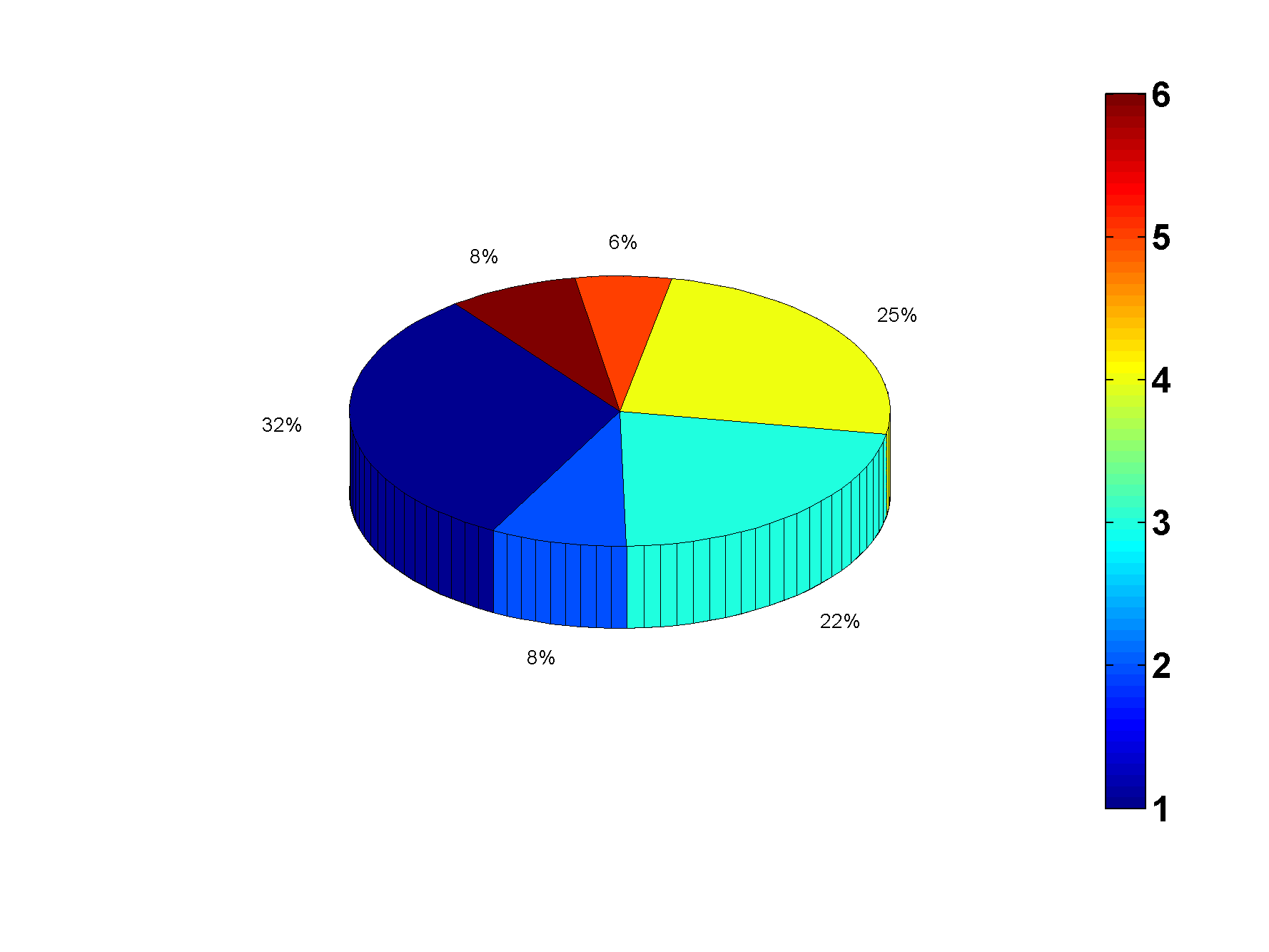}
\end{minipage}
}
\\
\subfloat[SNR=10]{
\begin{minipage}[t]{0.25\textwidth}
\centering
\includegraphics[width=1.0\textwidth]{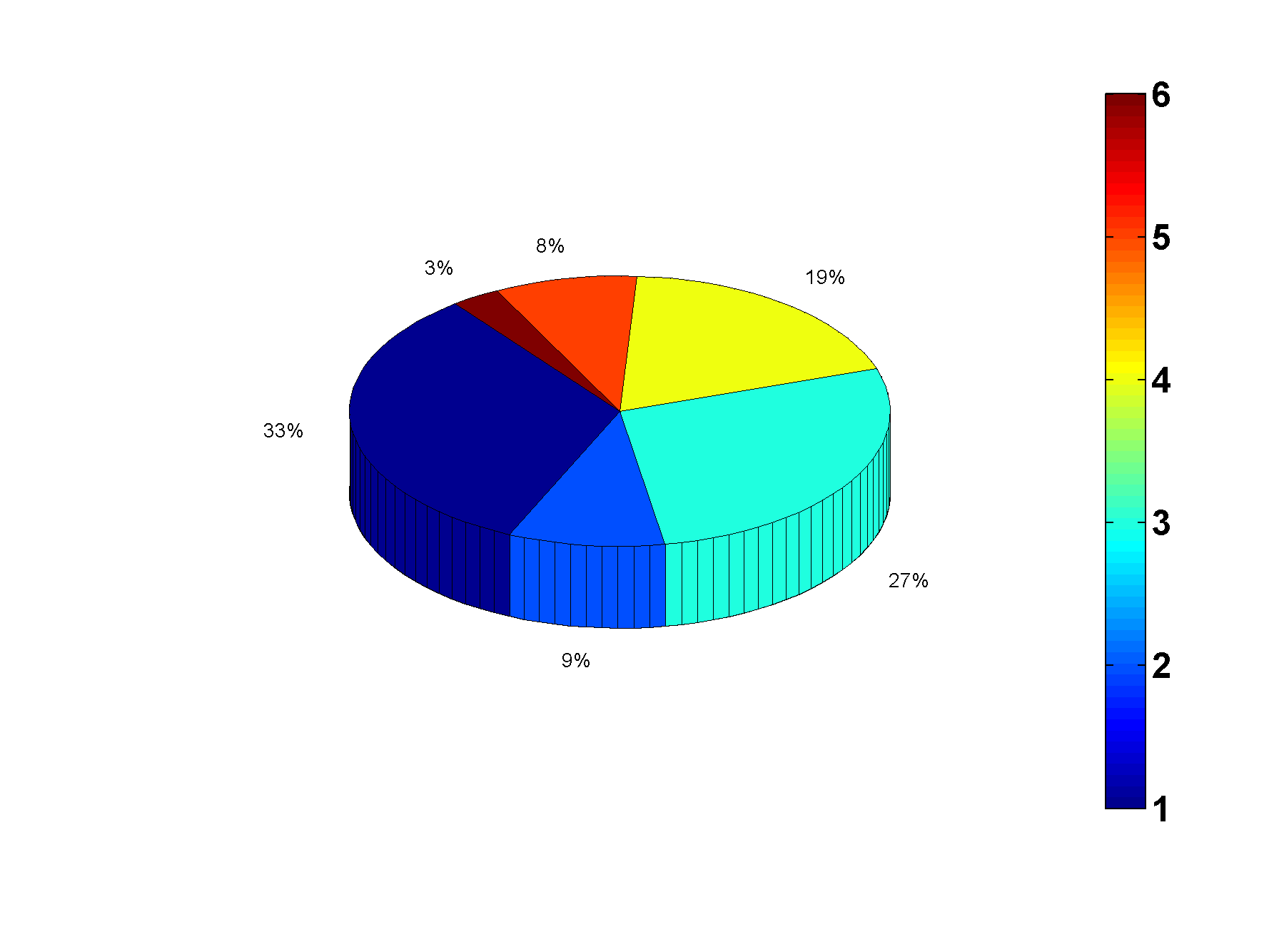}
\end{minipage}
}
\subfloat[SNR=10]{
\begin{minipage}[t]{0.25\textwidth}
\centering
\includegraphics[width=1.0\textwidth]{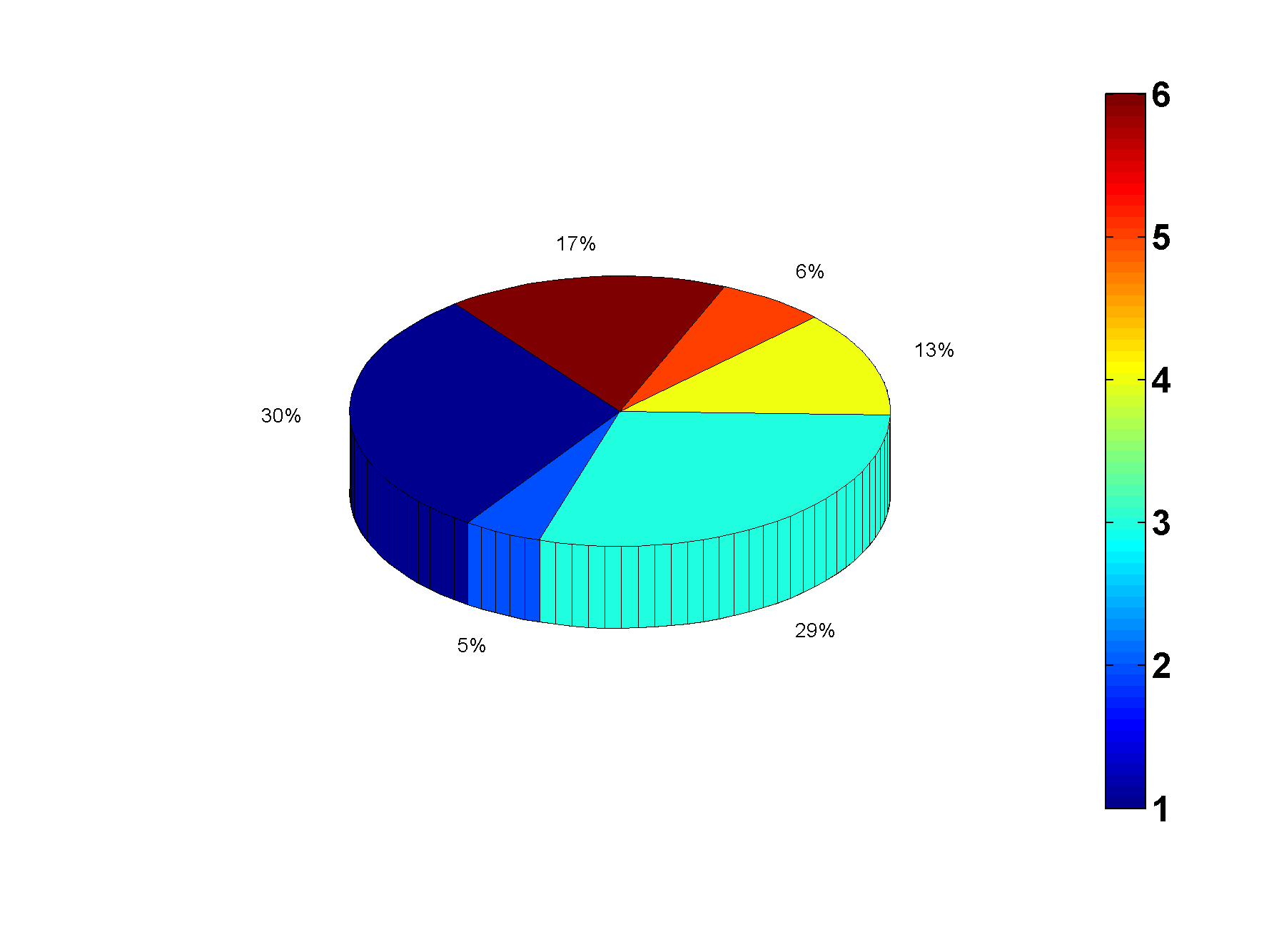}
\end{minipage}
}
\caption{ \label{fig:SNRpie} Pie charts of $X_k$ for different SNRs. (e) and (f) are for the same SNR with different noise realizations.
The color bar denotes the index of $X_k$. }
\end{figure}

Next, let us examine the role of the threshold $\mathcal{T}=\frac{1}{2}\max(X_k)$. In the previous simulations, we have six inner products $X_k,\;(k=1,\dots,6)$, each
corresponding to an SNR achieved by $H_i$. Since the detection criteria only depends on the ratio between the inner products, it is convenient to look at their pie charts. In Fig.~\ref{fig:SNRpie}, we show the pie charts for different SNRs, where the color bar represents the indices of the inner products. Take Fig.~\ref{fig:SNRpie} (a) for instance. The inner products $X_1,\,X_3,\,X_4$ contribute most part of the summation $\sum_{k=1}^6 X_k$, while $X_2,\,X_5,\,X_6$ are much smaller. According to the criteria we designed before, only $X_1,\,X_3,\,X_4$ are above the threshold. Therefore, we obtain the index $101100_2=44$ of the template, which most resembles the signal in the data. Similarly, Fig.~\ref{fig:SNRpie} (b)-(e) all successfully identify the correct template in case of different SNRs. Fig.~\ref{fig:SNRpie} (f) shows a failure case. According to the previous criteria, this pie chart gives a wrong index $101001_2=41$. In fact, even if one bit of the binary is wrongly determined, we end
up with a completely different template (and its corresponding parameters). This is also a main reason why the detection rate at low SNRs drops so quickly.

\emph{Improve the performance of the algorithm}--Now we discuss a simple and straightforward way to improve the performance of the algorithm by slightly
increasing the computational cost. Let us look at the failure case in Fig.~\ref{fig:SNRpie} (f) again. The largest inner product
is $X_1$, which contributes 30 percent of the entire SNR pie. The threshold, which was set to half of the largest inner product,
turns out to be $15$ percent. Therefore, among the six inner products, $X_1,\,X_3$ are significantly above the threshold, $X_2,\,X_5$
are significantly below, while $X_4,\,X_6$ are close to the threshold. In the end, the binary bits corresponding to $X_4$ and $X_6$ (i.e. the 4th and 6th bits)
were determined wrongly, which leads to a detection failure. However, the binary bits corresponding to $X_1,\,X_2,\,X_3$ and $X_5$
are correctly determined, and we are confident about that in the blind search. In fact, we are not so confident about the bits corresponding
to $X_4$ and $X_6$, since they are just slightly above or below the threshold. If we leave these two binary bits undetermined, we
end up with a binary number $101\textrm{y}0\textrm{y}_2$, where we have used $\textrm{y}$ to denote undetermined bits. It implies
that the true signal might match one of the four templates $101000_2=40$, $101100_2=44$, $101001_2=41$ and $101101_2=45$. By simply
calculating the inner products of the data and these four templates, we will know which one matches the true signal.

\begin{figure}
\centering
\includegraphics[width=0.5\textwidth]{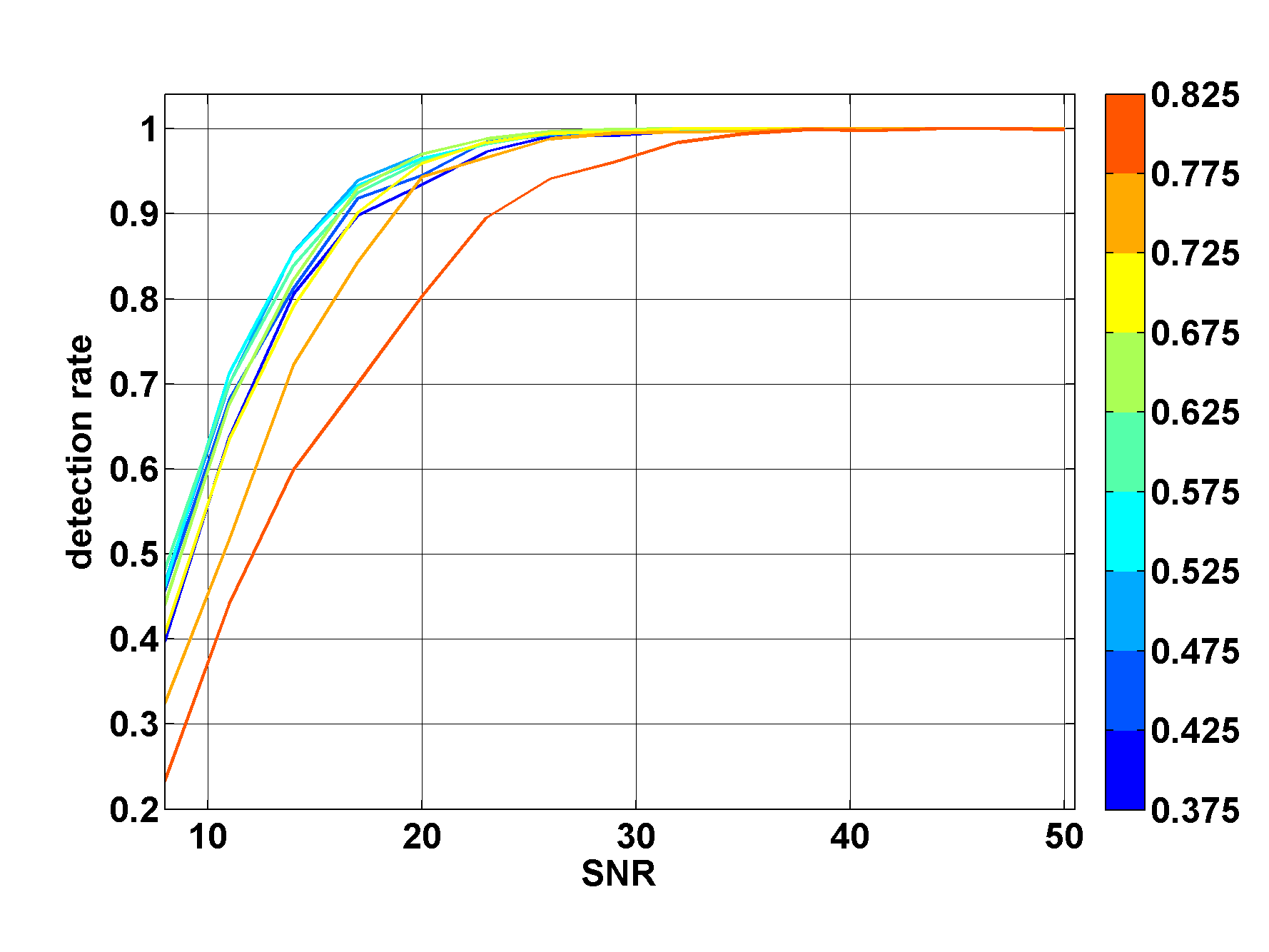}
\caption{ \label{fig:detectionRate2} Detection rate at different SNRs and thresholds. The color bar indicates the value of the coefficient
$c$. }
\end{figure}

Hence, we can modify the algorithm according to the above procedure. In the beginning, we calculate $X_k,\,(k=1,\dots,6)$ and
the threshold $\mathcal{T}= c\cdot\max(X_k)$. Then, we identify two $X_k$, which are closest to the threshold $\mathcal{T}$,
and leave two binary bits corresponding to these two $X_k$ undetermined. We determine other binary bits in the same way as before.
A binary number with two unknown bits is thus constructed. It corresponds to four original templates $h_i$. In the end, we
calculate the inner product between the data and these four templates, and detect the signal. Following this procedure, we
carried out a similar simulation as before. The detection rate is plotted in Fig.~\ref{fig:detectionRate2} with different combinations
of $c$ values and SNRs. Comparing with Fig.~\ref{fig:detectionRate}, the modified algorithm has significantly improved the
performance. The detection rate is increased at all SNRs. We also observe that $c=0.5$ is still the optimal choice. For the curve
$c=0.5$, the detection rate is 100\% above SNR 30 and 96\% at SNR=20. This strategy can be easily generalized by assigning a
probability to each binary bit according to $X_k$, hence obtaining the probability of each $h_i$ present in the data.
However, this is out of the scope of the current article. We will discuss it in the future work.

\emph{Conclusion and future work}--We have designed a novel algorithm for GW data analysis. Instead of using $2^N$ normal waveform templates, this new algorithm
uses only $N$ combinations of the original waveforms as the new templates. By calculating the inner products between these
$N$ new templates with the data and comparing these inner products with some threshold, we can construct a binary number with
$N$ bits. From this binary number, we can determine which normal template in the original template bank best matches the signal
in the data, without any reconstruction process. Therefore, this new algorithm can greatly reduce the computational cost in
certain circumstances. However, it requires relatively high SNRs. We have discussed a simple and straightforward way to improve
the performance of the algorithm. By leaving two most unconfident binary bits undetermined and calculating four additional inner
products, we can significantly improve the performance of the algorithm at low SNRs. The detection rate of the modified algorithm
is $100\%$ for 1000 different noise realizations for each SNR larger than 25. For SNR lower than 25, further improvements are
demanded. We reserve that for future work.

One possible way to improve the algorithm is to construct additional $H_k,\,(k=N+1,\dots)$ for auxiliary use, such as to determine
unconfident binary bits, to suppress the noise in $X_k$, etc. One can also set more sophisticated thresholds. We have used a threshold
only depending on the relative values between the inner products $X_k$ for simplicity. A threshold also depending on the absolute
values of the inner products would help, since the probability distribution of the random part of $X_k$ depends only on the absolute
SNRs.

We have only carried out simulations for a bank of nearly independent templates. In the future, we will do a simulation for an
entire template bank. The correlation between templates need also to be studied, since it could be used to reduce the noise
in the detection statistic.

%%%%%%%%%%
\begin{acknowledgements}
Y. W. was partially supported by DFG Grant No. SFB/TR 7 Gravitational Wave
Astronomy and DLR (Deutsches Zentrum f\"{u}r Luft- und Raumfahrt). Y. W.
also would like to thank the German Research Foundation for funding
the Cluster of Excellence QUEST-Center for Quantum Engineering and
Space-Time Research.
\end{acknowledgements}

%%%%%%%%%%%%%%%%%%%%%%%%%%%%%%%%%%%%%%%%%
\end{document}